\documentclass[print,twocolumn,superscriptaddress,preprintnumbers,amsmath,amssymb,epsfig,floatfix,prb]{revtex4-1}
\usepackage{graphicx}
\usepackage{dcolumn}
\usepackage{bm}
\usepackage{wrapfig}
\usepackage{amssymb}
\usepackage{soul}
\usepackage{amsmath}
\usepackage[breaklinks=true,colorlinks,linkcolor=blue,citecolor=green]{hyperref}
\def\be{\begin{equation}}
\def\ee{\end{equation}}
\def\e#1{\label{#1}\end{equation}}
\def\bea{\begin{eqnarray}}
\def\eea{\end{eqnarray}}
\def\ea#1{\label{#1}\end{eqnarray}}

\def\bem#1{\begin{mathletters}\label{#1}}
\def\eml{\end{mathletters}}

\def\ket#1{{|#1\rangle}}

\def\mean#1{{\langle#1\rangle}}
\def\4#1{{\boldsymbol{#1}}}
\def\8#1{{\widetilde{#1}}}
\def\bse{\begin{subequations}}
\def\ese{\end{subequations}}

\def\Rb87{$^{87}\text{Rb}$}
\def\0{\ket{0}}
\def\1{\ket{1}}

\newcommand{\Hres}{H_{\mathrm{res}}}
\newcommand{\Hdisp}{H_{\mathrm{disp}}}

\begin{document}
\title{Spin-bath polarization via disentanglement}
\author{D. D. Bhaktavatsala Rao}
\email{d.dasari@pi3.uni-stuttgart.de}
\affiliation{3. Physikalisches Institut, University of Stuttgart, Pfaffenwaldring 57, 70569 Stuttgart, Germany}
\author{Arnab Ghosh}
\affiliation{Department of Physics, Shanghai University, Shanghai 200444, People's Republic of China}
\affiliation{Department of Chemical Physics, Weizmann Institute of Science, Rehovot~7610001, Israel}
\author{David Gelbwaser-Klimovsky}
\affiliation{Department of Chemistry and Chemical Biology, Harvard University, Cambridge, MA 02138}
\author{Nir Bar-Gill}
\affiliation{Dept. of Applied Physics, Rachel and Selim School of Engineering, Hebrew University, Jerusalem 91904, Israel}
\affiliation{Racah Institute of Physics, Center for Nanoscience and Nanotechnology and Center for Quantum Information Science, The Hebrew University of Jerusalem, Jerusalem 91904, Israel}
\author{Gershon Kurizki}
\email{gershon.kurizki@weizmann.ac.il}
\affiliation{Department of Chemical Physics, Weizmann Institute of Science, Rehovot~7610001, Israel}
\affiliation{Department of Physics, Shanghai University, Shanghai 200444, People's Republic of China}


\date{\today}

\begin{abstract}
Spin bath polarization is the key to enhancing the sensitivity of quantum sensing and information processing. Significant effort has been invested in identifying the consequences of quantumness and its control for spin-bath polarization. Here, by contrast, we focus on the adverse role of quantum correlations (entanglement) in a spin bath that can impede its cooling in many realistic scenarios. We propose to remove this impediment by modified cooling schemes, incorporating probe-induced disentanglement via alternating, non-commuting probe-bath interactions, so as to suppress the buildup of quantum correlations in the bath. The resulting bath polarization is thereby exponentially enhanced. The underlying thermodynamic principles have far-reaching implications for quantum technological applications.
\end{abstract}

\maketitle

\noindent

The laws of thermodynamics determine the bounds on the efficiency, speed and minimal temperature achievable by cooling processes \cite{first}. Are there qualitative differences between these cooling bounds for quantum and classical systems? Such questions have come to the fore in recent years on both fundamental and practical grounds \cite{first}. A class of quantum systems where cooling bounds are highly important are spin baths in solids or liquids, as their cooling to low temperatures, alias hyperpolarization, \cite{slichter1990principles} would drastically boost the resolution of MRI or the probing sensitivity in NMR or quantum magnetometry \cite{jw}.

Here we address the almost inevitable role of intra-bath quantum correlations or entanglement that arise from the collective interactions of the spins with the probe that induces the cooling. We underscore the  extent to which such entanglement hinders the spin-bath polarization and propose to overcome this hindrance by disentangling the bath, as the core of a novel, highly effective cooling protocol. Such a protocol defies the prevalent current trend whereby quantum features give rise to "quantum supremacy" in thermodynamic processes \cite{first}.

In the quantum domain~\cite{Grinolds2013,isslerprl10,togan2011laser}, energy exchange between spins is maximized when they undergo resonant exchange (flip-flop), known as the Hartmann-Hahn (HH) effect~\cite{hartmann1962nuclear,slichter1990principles}. Under the HH conditions, widely used in NMR and MRI, a low-polarized (hot) spin can be cooled through polarization swapping with a higher-polarized (cold) spin. If the cold spin is a probe that is continuously polarized by an external source, i.e., if the probe entropy is nearly-instantly removed, this probe may be expected to fully polarize the entire spin bath, i.e., bring about hyperpolarization\cite{slichter1990principles}. This would occur if it were not for the quantum correlations (entanglement) that are ubiquitous in spin baths and often have a very long decay time. Studies of spin-bath cooling by a spin probe have shown that the stationary states of such a bath may be highly entangled, and the corresponding polarization of each spin in the bath is much lower than that of the spin bath \cite{ciracprb}. The reason is the collective coupling to the spin probe that causes the spin-bath to evolve into \textit{disjoint manifolds} of the collective spin observables even when the couplings of the individual spins to the probe vary significantly (Fig. 1). These manifolds (subspaces) remain invariant under resonant exchange with the probe \cite{Niedenzu2018} and are a bottleneck for cooling, since they block heat and entropy removal from each subspace via resonant swap with the probe. Earlier attempts to break the spin collectivity have resorted either to wavefunction modulation of the probe, so as to controllably vary its interactions with the individual spins in the bath \cite{wavefuncmod}, or to spin-spin interactions among  bath spins that lift their degeneracy \cite{dipolarlukin}. The former control is not applicable to dipolar-coupled spin systems, e.g. to NV centers coupled to nuclear spin baths.The latter mechanism  was shown to enhance the spin polarization in the bath, but only up to $30\%$ \cite{dipolarlukin}. Here we put forward  a general prescription for  enhancing spin polarization in the bath nearly to $100\%$ via collectivity destruction, i.e. disentanglement.  We quantitatively  show that this approach is applicable to both NV centers in nuclear spin baths and hyperfine-coupled  quantum-dot systems.  

Our goal is to fully polarize all spins i.e., bring them to their ground state, irrespective of the spin-probe couplings variability. One option is to have strong inhomogeneous broadening so as to assign  to a different resonance to each spin~\cite{ciracprb}. This would, however, preclude the HH resonant swap and thus hinder cooling. We here advocate \textit{dynamic symmetry-breaking control} that repeatedly destroys the correlations (entanglement) induced by the collective coupling. We show that it is possible to nearly fully polarize the spins in the bath to their ground state, efficiently and rapidly, by alternating resonant and off-resonant (dispersive) coupling of the probe to the bath, thereby inducing in turn flip-flop and disentangling interactions in the spin-bath. 
We then show that spin-spin interactions may facilitate spin polarization provided they are comparable in strength to the individual spin-probe couplings. 
%

\par 

Since disentanglement which is  a form of decoherence, tends to increase the state entropy it is a counter-intuitive means of  cooling down an unpolarized spin ensemble, e.g., a nuclear spin-ensemble cooled by an electron spin of a defect center in a diamond \cite{TaylorNAT08} or in a semiconductor quantum-dot \cite{Maletinsky2009,Hildman14,Gunter16} It opens a new vista into the effects of quantum correlations on entropy changes in an open many-body system and their thermalization ~\cite{Erez2008,David15,Gogolin2016,Gring1318,AlvarezPRL10}.

\begin{figure}[h]
\includegraphics[width=1.0\linewidth]{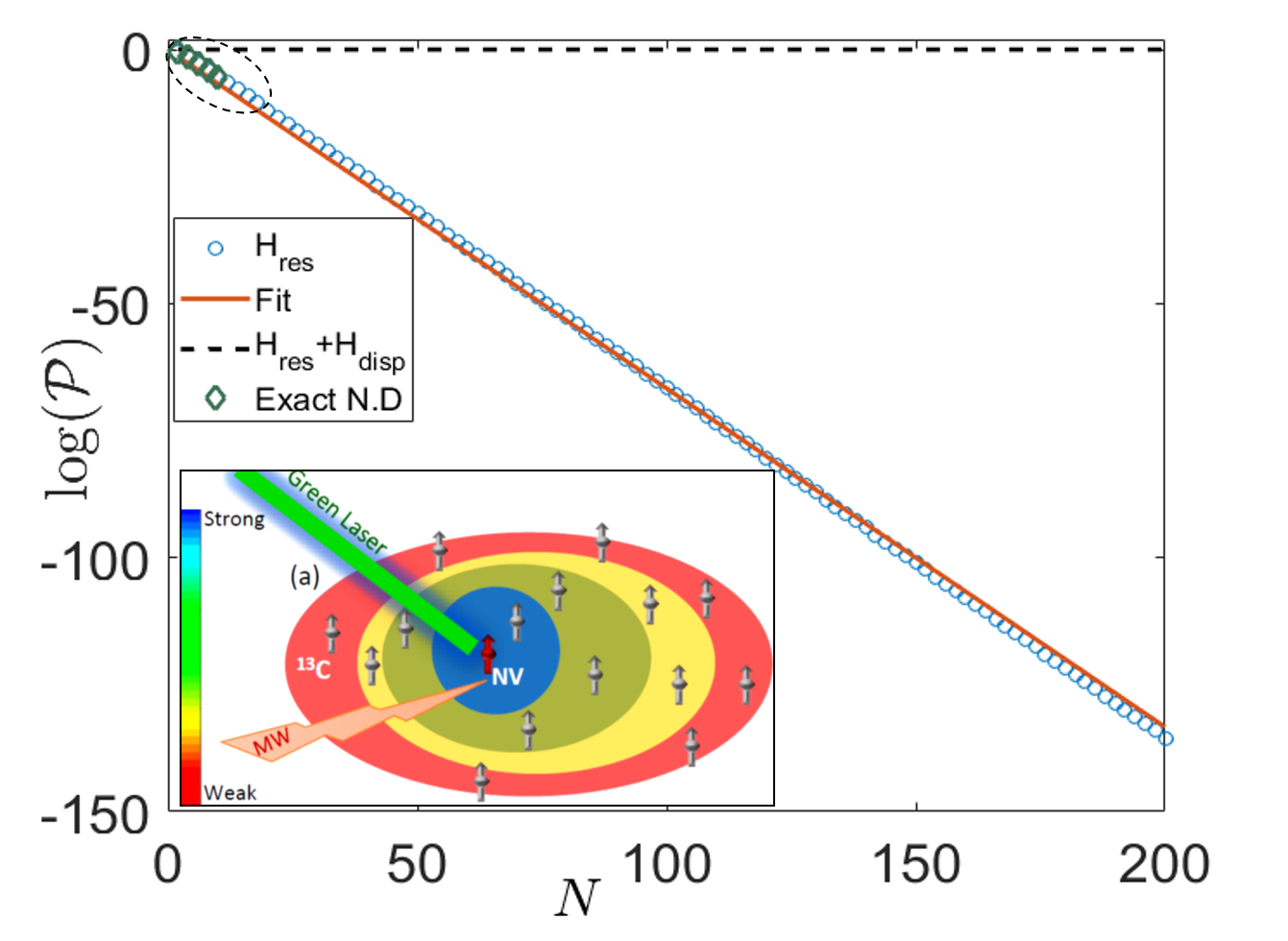}
\caption{\label{fig:intro}
The bath purity ($\log{\mathcal{P}}$) exhibits an exponential decay with the bath size $N$. The analytical formula (Eq. 2) is verified by numerical diagonalization (N. D) for $N \le 10$ (green diamonds). Inset: A central spin (NV center) interacting with a spin-bath (consisting of $^{13}$C nuclear spins). The central spin is optically pumped (O.P) by a green laser to dispose of its entropy and coherently manipulated by microwave (MW) pulses. The color bar indicates the strength of coupling between the NV and bath spins.  }
\end{figure}

\par
\begin{figure*}[tbhp]
\centering
\includegraphics[width=1.0\linewidth]{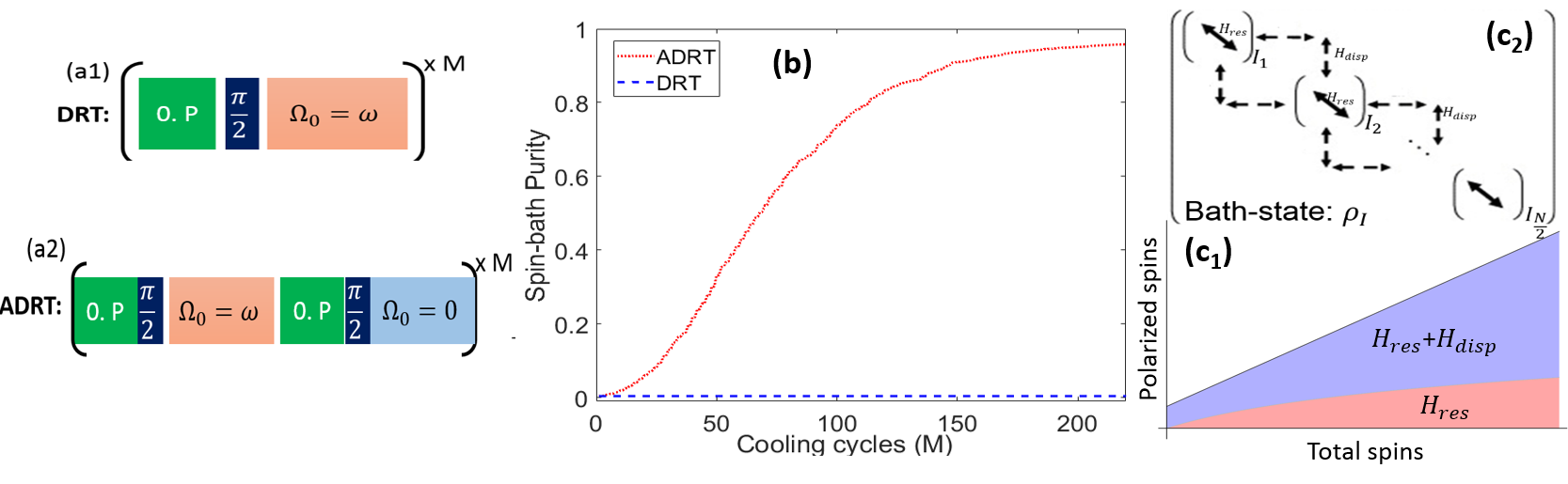}
\caption{\label{fig2} ({\bf a})  (a1) The standard M-fold pulse sequence for polarizing the nuclear spins implemented by the Hartman-Hahn (HH) double resonance transfer (DRT), which leads to the formation of invariant subspaces that hinder further polarization exchange. (a2) The proposed pulse sequence, consisting of alternating dispersive and resonant transfer (ADRT), which destroys the symmetry imposed by the conservation of the  total spin $I$, by modulating the probe spin in and out of resonance. ({\bf{b}}) Drastically different degrees of spin-bath purity achievable by DRT (blue dashed-line) and ADRT (red thin dashed-line) as a function of the cycle number $M$,  for $N$-nuclear spins that are dipole-coupled to an electron probe spin of an NV-center in diamond. Here we analyze the case of $N=8$, nuclear spins, dipolar coupled to the electron spin. The couplings are weak with respect to the nuclear Larmor frequency $\omega_L$ and are chosen in the range $0 <\frac{|\vec{g}|}{\omega_L} < 10^{-2}$ for random locations of the spins in the $x-z$ plane. (${\bf{c_1}}$) Spin-bath polarization under alternating $H_{res}$ and $H_{disp}$ is much higher than under $H_{res}$ alone. (${\bf{c_2}}$) Schematic evolution of a block-diagonal spin-bath-state. Each block (manifold) is labeled by its total spin $I_j ~(j = 1\cdots N/2)$. Under alternating resonant ($H_{res}$ causing intra-block polarization) and off-resonant/dispersive ($H_{disp}$ causing dephasing), the state undergoes inter-block mixing.}
\end{figure*} 

\noindent
{\bf Concept:} We first consider a scenario wherein a spin probe is coupled to $N$ identical non-interacting spins. On resonance, we then obtain an energy-exchange flip-flop Hamiltonian
\be
H  = \sum_k g_k (S^+  I^-_k + S^-  I^+_k).\label{eq:H}
\ee
For identical couplings, i.e., $g_k = g$, we find that the total spin $I$ of the ensemble remains conserved and the Hamiltonian assumes a block-diagonal form so that the dynamics takes place independently in each block (manifold) with fixed $I$. Consequently, no population exchange can take place among manifolds corresponding to collective-spin subspaces with different $I$. 
Resonant exchange of the spin ensemble via Hamiltonian \eqref{eq:H} with a frequently polarizable spin probe $S$ allows each manifold to be polarized (purified) independently. As mixing among spin-manifolds with different $I$ is not allowed by total spin conservation, the total purity of the spin-bath is  suppressed, even though the spin probe is able to transfer its entire polarization to the spin-bath. The resulting purity $\mathcal{P}$ of the spin-bath under such resonant exchange decays exponentially with $N$ (Fig. 1):
\be
\mathcal{P} \equiv {\rm Tr}[\rho^2_I] = \sum_{I = 0}^{N/2} \lambda_I \left(\frac{2I+1}{2^N}\right)^2 \sim {\rm e}^{-2N/3},
\ee\label{eq:Purity}
where $\lambda_I = {}^N C_{N/2-I} \frac{2I+1}{N/2+I+1}$ is the multiplicity of the different $I$-manifolds. Such suppressed purification of the bath holds even \textit{when the couplings $g_k$ are different} (see SI).

By contrast,\textit{disentanglement} of the bath spins leads to  population exchange among total-$I$ subspaces, thereby increasing the mixedness in each sector. Through repeated destruction of the quantum correlations in the spin-bath followed by heat exchange of the probe with a very cold (vacuum) bath, we can effectively pump all the population to the highest spin subspace $I=N/2$ and only to states with $\mean{I_z} = \pm N/2$. These are the only disentangled eigenstates of the total-spin operator  $I$, corresponding to a fully polarized (purified) spin-bath. 
Thus, counter-intuitively, exponential enhancement of $\mathcal{P}$ as a function of $N$ is achievable by \textit{erasing} rather than generating quantum correlations. 

Our approach is quite generally applicable to solid-state spin baths coupled to a spin probe ~\cite{Staudacher2013,Mamin2013,McGuinness2011,Grinolds2013,kucskonat13}. For the sake of concreteness, we discuss the NV center case in what follows.

The NV spin-probe (central spin) is a two-level system (TLS) at frequency $\omega_0$  or driven by a microwave (MW) field at resonance frequency $\omega_\mathrm{0}$ with the Rabi-frequency $\Omega_0$ according to the rotating-frame Hamiltonian (see SI)
\begin{equation}
H_{0}(t)=\frac{1}{2}\omega_{0}S_{z}+ \Omega_0\bigl(S^{-}e^{i\omega_\mathrm{0} t}+S^{+}e^{-i\omega_\mathrm{0} t}\bigr) +  \omega_\mathrm{L} \sum_k I^z_k ,\label{Ham}
\end{equation}
where $S_z$, $S^{\pm}$ are, respectively, the inversion, raising and lowering  Pauli spin operators. The adjacent nuclear  spins labeled by `$k$', with energy-splitting $\omega_L$, are viewed as a hot bath to be cooled down.
%

By restricting the bath-spin operators $\vec{I}_k$ to the $x-z$ plane, i.e., setting $\vec{g}_k\cdot\vec{I}_k =g^x_k I^x_k + g^z_kI^z_k$, and transforming to the interaction picture, the total Hamiltonian simplifies to \cite{jelezkoprl}

\begin{eqnarray}
\tilde{H} &=& \sum_k \big [g^x_k\left({\rm e} ^{-i(\Omega_0 - \omega_\mathrm{L})t}S^+I^-_k + {\rm e} ^{-i(\Omega_0 + \omega_\mathrm{L})t}S^+I^+_k + h.c\right) \nonumber\\
&+& g^z_k\left({\rm e} ^{-i\Omega_0 t}S^ + + h.c\right) I^z_k \big].\label{H-RWA}
\end{eqnarray}
where $g^x_k$ and $g^z_k$ are the spin-probe couplings components.
The HH-condition for polarization transfer (swap) between the probe and the bath is fulfilled in the dressed basis by setting $\Omega_0 = \omega_\mathrm{L}$, known as the double-resonance transfer (DRT) (Fig.~\ref{fig2})~\cite{jelezkoprl}. By contrast, for dispersive (off-resonant) coupling between the probe and the bath, $\Omega_0 = 0$, probe-induced energy shifts of the bath spins occur without any polarization transfer. Thus, we can have two contrasting regimes for the combined probe-bath dynamics governed by Eq.~\eqref{H-RWA}
\bea
\Omega_0 = \omega_d &\implies& ~\tilde{H} \equiv H_\mathrm{res} = \sum_kg^x_k(S^+ I^-_k +S^- I^+_k ), \label{H_res} \\
\Omega_0 = 0 &\implies&~ \tilde{H}  \equiv H_\mathrm{disp} = S^z \sum_kg^z_k I^z_k \label{H_disp}.
\eea

\par

We show here that alternating evolutions governed by Eqs.~\eqref{H_res} or~\eqref{H_disp}, respectively, which 
we dub alternating dispersive-resonant transfer (ADRT) are required to maximize the polarization of the spin bath, as opposed to DRT that only employs resonant exchange (Eq.~\eqref{H_res}).

\par

In addition to the MW control above, the probe ($S$) spin is subject to optical (laser) pumping: a transition to an electronic excited state followed by entropy dumping via coupling to an electromagnetic (EM) bath. This process results in the rapid relaxation of the probe spin to its ground state, since at optical frequencies the EM-bath is effectively empty, i.e., has Boltzmann factor $\beta \omega_\mathrm{opt} \rightarrow \infty$, $\beta$ being the inverse temperature. 

\par 

Let us repeatedly reset the probe to its ground state by optical pumping, and let the spin dynamics be induced alternately by Eqs.~\eqref{H_res} or~\eqref{H_disp} between consecutive resettings. These alternating interactions realized by modulating the resonance frequency of the probe yield the ADRT (Fig.~\ref{fig2}a)


Due to the repeated resetting of the probe spin by optical pumping, its correlations to the spin bath are erased, leaving the total state $S+B$ uncorrelated at any time $t$, i.e., $\rho(t) = \rho_S(t)\otimes\rho_\mathrm{B}(t)$, where $\rho_\mathrm{B}$ denotes the spin-bath state. Hence, the probe spin can be traced out to obtain an effective \textit{non-Markovian master equation} (ME) for the spin-bath state $\rho_\mathrm{B}(t)$ to second order in the couplings $\vec{g}$ (see SI).  During the alternating driving of the bath by $H_\mathrm{res}$ and $H_\mathrm{disp}$, the ME dissipators are constructed from the collective spin operators of the bath associated with Eqs. ~\eqref{H_res} and ~\eqref{H_disp} respectively. The $S^+I_k^-, S^-I_k^+$ interactions of the bath spins with the probe in Eq.~\eqref{H_res}, lead to relaxation (polarization) at the respective rates $\Gamma_+(t)$ and $\Gamma_- (t)$, whereas the interaction in Eq.~\eqref{H_disp} yields decoherence at the rate $\Gamma_z(t)$.

\begin{figure*}[htbp]
\centering
\includegraphics[width=1.0\linewidth]{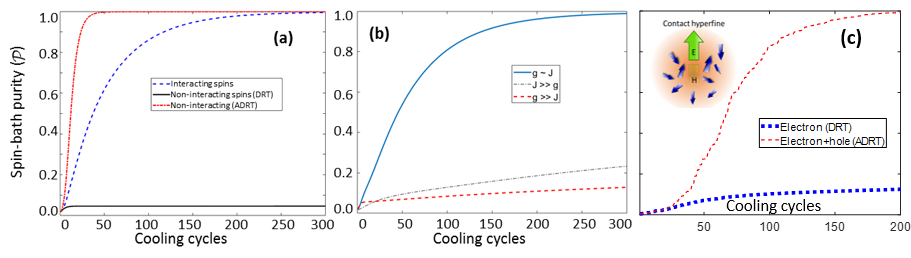}
\caption{(a) Comparison of spin bath purification for (i) interacting spins (blue-dashed), (ii) non-interacting spins under ADRT (red-dotted), and (iii) non-interacting spins under DRT (black solid), obtained numerically are shown. (b) The role of spin-spin interactions among the nuclear spins in achieving the maximal purity, is numerically shown to be similar to that achieved by ADRT. We consider a bath of six spins, where the $S-I$ and $I-I$ interactions [see SI] are given by $g_k = g\lbrace 1,1,1,1,1,1 \rbrace$, and nearest-neighbor coupling among the bath spins with $J_{ij} = \alpha \lbrace 0.8,1.0,1.2,1.3,2.1 \rbrace$. The strength of coupling $\alpha$ in the three cases shown are respectively (denoting $J_{ij} \sim J$) $\alpha = 1$ ($g \sim J$), $\alpha = 0.1$ ($g \gg J$), $\alpha = 10$ ($g \ll J)$, and only DRT interaction between the bath and the probe. With the inhomogeneous coupling chosen among the bath spins, the unique common eigenstate of the Hamiltonian is $\ket{\downarrow\downarrow\cdots\downarrow}$ i.e., a fully polarized bath, whose polarization is similar to that of the central spin.(c) Numerical simulations of similar protocols (shown in Fig. 2) applied to the case of nuclear spins that are hyperfine-coupled to either electron-or hole-spins in a semiconductor quantum dot. The protocol analogous to ADRT is realized by modulating optical pulses,thereby causing nuclear spins to evolve with random phases through hyperfine coupling to both electron-and hole-spins. The analog of DRT is realized when the nuclear spins evolve only under coupling to the electron spin. The analog of ADRT shows great improvement over the analog of DRT.} \label{fig:3}
\end{figure*}

\par
Under DRT, the total spin of the ensemble remains conserved, even though the individual spin couplings $g^x_k, g^z_k$ may be different. Hence, no population exchange can take place among various manifolds labeled by the total spin $I$, that are associated with the collective-spin angular-momentum operators, 
\begin{eqnarray}
\tilde{I}^\pm_{\mathrm{res}}= \frac{1}{{g_\perp}}\sum_k g^x_k I^\pm_k;  \quad g_\perp=\sqrt{\sum_i (g^x_i)^2}.
\end{eqnarray}

By contrast, upon switching  $H_\mathrm{res}$~\eqref{H_res} off and $H_\mathrm{disp}$~\eqref{H_disp} on, the conserved operator becomes,
\begin{eqnarray}
\tilde{I}^z_{\mathrm{disp}} = \frac{1}{{g_\parallel}} \sum_k g^z_k I^z_k, \quad g_\parallel=\sqrt{\sum_i (g^z_i)^2},
\end{eqnarray}
 instead of total spin. The dynamics that is alternately governed by either $\Hres$ or $\Hdisp$ then leads to population mixing among various spin manifolds conserved by either $\tilde{I}^z_\mathrm{disp}$ or $\tilde{I}^{\pm}_\mathrm{res}$, since the respective invariant manifolds are \textit{different}. Consequently, the invariant correlation-induced manifolds in the spin bath are periodically destroyed (erased) under ADRT because the alternating dominance of $\Gamma_{\pm}(t)$ or $\Gamma_z(t)$ effectively projects $\rho_B$ onto the collective basis-state of $\tilde{I}^\pm_\mathrm{res}$ or $\tilde{I}^z_\mathrm{disp}$ respectively. If the couplings $g^x_k$ were equal, the alternation and likewise $g^z_k$, of Hamiltonians~\eqref{H_res} and~\eqref{H_disp} would merely rotate the collective states but not destroy their collectivity. 
Yet, because of the inequivalence of the two collective bases, the ADRT modulation of the spin-probe level distance by $\Omega_0(t)$ between $\omega_d$ and $0$ allows unrestricted population mixing among all bath manifolds, followed by the cooling down of the spin-bath via  HH resonant transfer and spin-probe entropy dumping.  Since the cold (EM) bath to which the  probe spin is coupled satisfies $\beta \omega_\mathrm{opt} \rightarrow \infty$, a steady constant rate of polarizing the probe spin is reached, the cooling rates of the bath also reach their asymptotic limits, i.e., $\Gamma_+  \rightarrow g_\perp, ~\Gamma_z \rightarrow g_\parallel$, whereas $\Gamma_- \rightarrow 0$ (see SI). The ME then yields 
decay to the ground state of the spin-bath. 

\par
The superiority of ADRT over DRT is conspicuous (Fig.~2 (b)). The effect of two competing alternating angular momentum bases that are formed by the probe-bath alternating resonant and dispersive interactions can purify the spin-bath to their common eigenstate $\ket{N/2;\pm N/2}$ i.e., to a fully polarized state. Such competing bases may also be obtained when considering an interacting spin system. Let us consider a scenario where a central spin is uniformly coupled to a spin-bath, wherein the spins interacts with each other. 
In the rotating frame with $\omega_0 = \omega_L$, we then obtain the resonant exchange Hamiltonian, which now ncludes the intra-bath spin-spin interactions i.e.,
\be
H= S^+\sum_k g_k I^-_k + \sum_{i,j} J_{i,j}I^+_iI^-_j + h.c.
\ee
For either of the terms in this Hamiltonian, the fully polarized state $\ket{N/2;\pm N/2}$ is an eigenstate. The symmetrized basis for the evolution of the spin-bath depends on the relative strengths of $g_k$ and $J_{ij}$. Polarization transfer is incomplete when either of these terms dominate the dynamics, since the basis they consists of collective states, that compete with each other, so that their only common eigenstate is the ground state $\ket{N/2, -N/2}$ (see SI). This anticipation is confirmed by exact numerical diagonalization of the above Hamiltonian in Fig. 3. The role of spin-spin interactions among the nuclear spins in the bath may thus be similar to that of ADRT, namely, spin-spin interactions enable full purification.  As the interactions lead to the formation of different total-$I$ subspaces, the symmetry imposed by DRT is invariably broken. In comparison with ADRT, the purity gain of interacting spins is slow, but eventually saturates to the fully polarized state of the bath.

\noindent
{\bf Conclusions:}The fact that quantum correlations (entanglement) induced among the bath spins by their coupling to a spin probe may live long leads to a strong deviation  of the spin bath from the initial thermal state as it resonantly interacts with the spin probe giving rise to complex many-body dynamics. The spins are driven into invariant manifolds that inhibit any further mixing among these manifolds due to their coupling to the spin probe, thus preventing the bath cooling. Hence, only by frequently obliterating the correlations among the bath spins may one achieve maximal cooling of the bath.  To this end, we have introduced the unconventional sequence we dub alternate dispersive-resonant transfer (ADRT) wherein disentanglement of the spin bath (by modulating the spin-probe energy) alternates with the common flip-flop regime (the HH double resonance transfer - DRT). While DRT  leads to a spin-bath polarization (purity), that decreases  exponentially with the size of the bath, ADRT can nearly-fully polarize it. The present insights hold for interacting spin baths, where spin-spin interactions have been shown to yield effects similar to ADRT.

\par

\par

Controlled destruction of correlations (disentanglement) among quantum systems can be a key to understanding the thermalization of quantum systems coupled to finite baths, and the observed equilibration of quantum systems. The proposed strategy may be highly useful in reducing the noise produced by the surrounding electronic or nuclear spin bath on a probe spin by  polarizing these spin baths in probed samples. Thereby, we may drastically enhance the performance of sensing, magnetic imaging and spectroscopy, metrology and quantum information processing schemes~\cite{TaylorNAT08,grinolds14,kucskonat13,neumannnanolet13,toylipnas13,mittermaierscu06,
smithpnas12,Staudacher2013,Mamin2013,Grinolds2013,rondin13,wrachtrupefield}.

\section*{Acknowledgements}
G. K. acknowledges support by the ISF, EU (PATHOS, FET Open) and SAERI.  D. D, and G. K jointly acknowledge the DFG support through the project FOR 2724. Bar-Gill acknowledges the support from the European Union (ERC StG, MetaboliQs), the CIFAR Azrieli Global Scholars program, the Ministry of Science and Technology, Israel, and the Israel Science Foundation (Grant No. 750/14).


%


\end{document}